\newlength\smallfigwidth
\newlength\figwidth
\documentclass[aps,twocolumn,floatfix,amsmath,amssymb,showkeys,letterpaper]{revtex4}
\usepackage{amsmath}

\newcommand{\be}{\begin{equation}}
\newcommand{\ee}{\end{equation}}
\newcommand{\bn}{\begin{eqnarray}}
\newcommand{\en}{\end{eqnarray}}
\newcommand{\ii}{{\rm i}}

\usepackage{graphicx}
\usepackage{hyperref}
\usepackage{bm}

\begin{document}

\title{Metastability and dynamic modes in magnetic island chains}
 
  \author{G.\ M.\  Wysin}
  \email{wysin@phys.ksu.edu}
  \homepage{http://www.phys.ksu.edu/personal/wysin}
  \affiliation{Department of Physics, Kansas State University, Manhattan, KS 66506-2601}

\date{October 24, 2021}
\begin{abstract}
{The uniform states of a model for one-dimensional chains of thin magnetic islands on a nonmagnetic 
substrate coupled via dipolar interactions are described here. 
Magnetic islands oriented with their long axes perpendicular to the chain direction are assumed, 
whose shape anisotropy imposes a preference for the dipoles to point perpendicular to the chain.
The competition between anisotropy and dipolar interactions leads to three types of uniform states 
of distinctly different symmetries,  including metastable transverse or remanent states, transverse
antiferromagnetic states, and longitudinal states where all dipoles align with the chain direction.
%
%
The stability limits and normal modes of oscillation are found for all three types of states, even including
infinite range dipole interactions.
The normal mode frequencies are shown to be determined from the eigenvalues of the stability problem.
%
}
\end{abstract}
\maketitle

\section{Arrays of magnetic islands}
\label{intro}
Artificial spin lattices can be fabricated from thin elongated magnetic islands arranged on a 
nonmagnetic substrate, such as in one-dimensional (1D) artificial spin chains \cite{Nguyen17,Ostman18,Cisternas21} and 
two-dimensional (2D) artificial spin ice systems\cite{skjaervo19}.
%
%
The competition of strong shape anisotropy with dipolar interactions in artificial spin ice with two-state 
Ising-like \cite{Ising25,Onsager44} dipoles leads to 
a ground state that follows an ice rule \cite{Wang06,Nisoli13,Morgan11}, 
with antiferromagnetic order in square artificial spin ice.
There are also metastable excited remanent states of nonzero average magnetic moment, that 
result from the application of an applied magnetic field which is slowly turned off.
The different low energy ground and remanent states possess distinctive modes of oscillation relative to those 
states \cite{Gliga+13,Jung+16,Iacocca+16,Lasnier+20}, that can serve as signatures \cite{Arroo+19,Arora+Das21}
of those states.

In particular for 1D artificial spin systems, \"Ostman \textit{et al.} \cite{Ostman18} fabricated and
analyzed chains of mesoscopic magnetic islands with strong shape anisotropy (due to a high aspect ratio), 
which causes the effective spins to behave as Ising-like.
Nguyen {\it et al.} \cite{Nguyen17} considered how slight changes from 1D to quasi-2D lattice structure can 
affect the lowest states of artificial Ising spins, especially due to the geometric dependence of the dipole 
interactions.
Cisternas {\it et al.} \cite{Cisternas21} considered a model for a chain of a few coupled {\sc XY} magnetic 
dipoles with rotational inertia (magnetic charges on dumbbells with one angular coordinate) and the 
stability of its dynamic solutions.
Alternatively, we consider a model for a chain of artificial spins with easy-plane shape anisotropy
and weak uniaxial shape anisotropy within the easy plane,  such that their behavior is not Ising-like, 
but rather, more closely described by three-component Heisenberg-like \cite{Heisenberg28,Jiles91} dipoles. 
This model applies to a set of thin and somewhat elongated magnetic islands fabricated on a nonmagnetic substrate.
The possible stable and metastable uniform states are analyzed for varying uniaxial anisotropy strength 
$K_1$ of the individual islands, relative to the effective strength of nearest neighbor dipole interactions, $D$.
For fixed island shapes and center-to-center separations $a$, the energy constants $K_1$, $D$ and an easy-plane
anisotropy strength $K_3$ are determined by the thickness and aspect ratio of the islands \cite{Wysin+12}.
However,  $K_1$ and $D$ might be modified post-fabrication by pressure or elastic strain \cite{Edberg21},
or other yet to be found methods.
%

\begin{figure}
\includegraphics[width=\figwidth,angle=0]{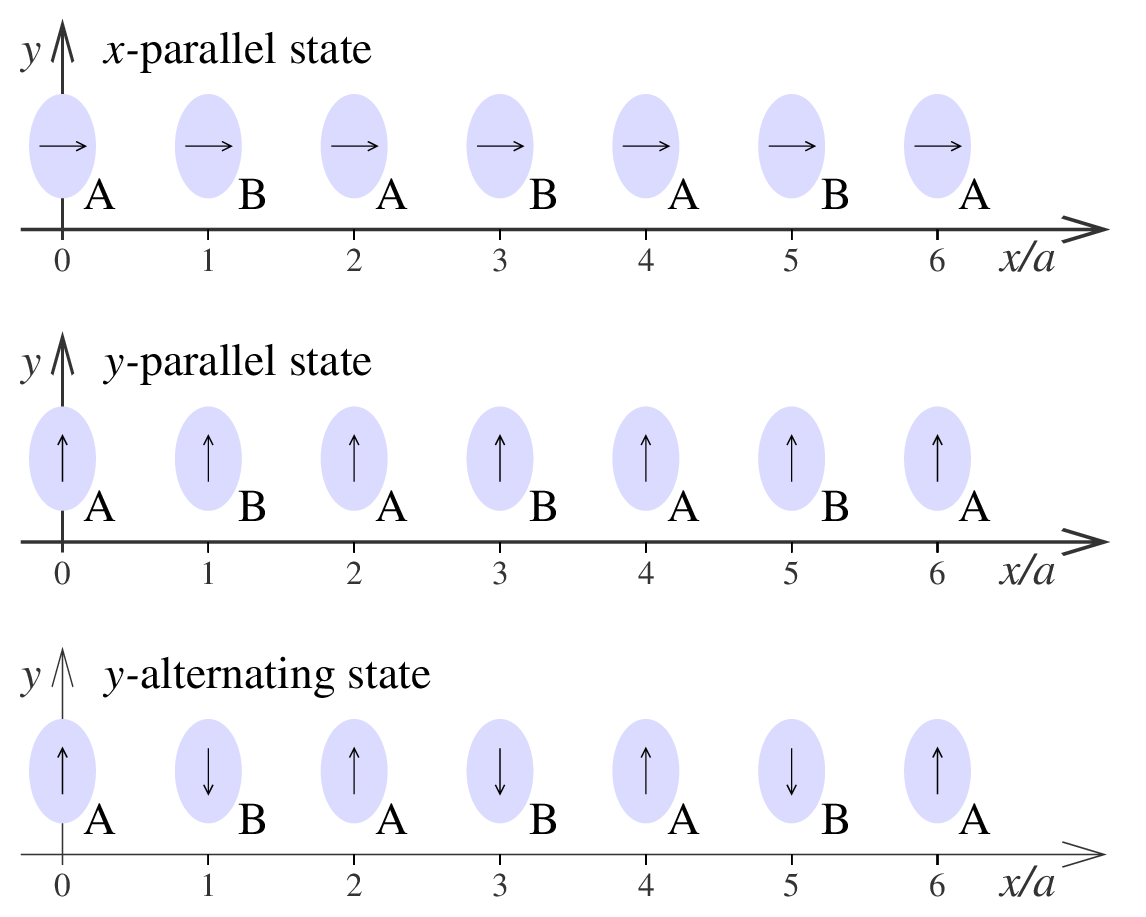}
\caption{\label{1d-islands} The three uniform states of a chain of elongated magnetic islands separated by 
lattice constant $a$, with long axes perpendicular to the chain direction.}
\end{figure}
%
%
The chain is taken along the $x$-direction, with elongated magnetic islands whose long axes are arranged perpendicular 
(along $y$) to the chain, see Fig.\ \ref{1d-islands}.
This is inspired by the geometry of a row of islands in artificial square lattice spin ice, however, 
the physics is different due to the one-dimensionality and the geometrical dependence of the dipolar 
interactions. 
The most interesting effects take place for relatively weak uniaxial anisotropy $K_1$ relative to $D$,
that would correspond to slightly elongated islands, as shape anisotropy is the result of
unequal demagnetization effects along the short and long axes of the islands. 
The Heisenberg-like island dipoles are assumed to be of fixed magnitude and able to point in any direction, but 
have energetic {\em preferences} for staying in the plane of the substrate, and for aligning with the islands' long axes. 

With long axes perpendicular to the chain, the islands' dipole moments will tend to point 
perpendicular to the chain direction to minimize the shape anisotropy energy. 
Then nearest neighbor dipolar interactions will impose a transverse alternating or antiferromagnetic  
(AFM) state (neighboring dipoles along $\pm \hat{y}$) for simultaneously minimizing the dipolar energy.
This AFM $y$-alternating state, depicted in Fig.\ \ref{1d-islands}, is reminiscent of a row in a ground state of 
square lattice artificial spin ice. 


Other possible uniform states are depicted in Fig.\ \ref{1d-islands}.
If a uniform magnetic field is applied perpendicular to the chain, and then turned off, it can leave 
the system in a corresponding metastable remanent or $y$-parallel state, where the dipoles either all 
point along $+\hat{y}$ or  all along $-\hat{y}$, with ferromagnetic order transverse to the chain.  
A uniform magnetic field applied parallel to the chain, and then turned off, could leave all 
the dipoles uniformly aligned with the chain direction, in an $x$-parallel state, with
ferromagnetic order longitudinally along the chain, but 
only if dipolar interactions dominate over the shape anisotropy ($D$ larger than $K_1$).
Nonuniform excited states, such as single dipole reversals from any of the uniform states, 
are not considered here.

The calculations here determine the static structure, the conditions for stability, and the 
linearized normal modes of oscillation for all three uniform states.
The stability analysis gives eigenvalues that are shown to connect directly to the 
dynamic mode frequencies. 
The description here is initially developed for a nearest neighbor model, 
and then extended to include infinite range dipolar interactions for an infinite chain.
The analysis is based on classical undamped spin dynamics, treating each island as a single macro-magnetic 
moment in the macrospin approximation used previously by Wysin {\it et al.}\cite{Wysin+13,Wysin+15} and
by Iacocca {\it et al.}\cite{Iacocca+16}.
While the macrospin assumption is certainly an approximation for a real-life magnetic island structure, 
the calculations should give some reasonable representation of what could happen in this type of 
engineered magnetic array, especially for adequately spaced islands where the neighbors' dipolar 
fields are quasi-uniform within an island \cite{Shevchenko17}.

\section{The magnetic island model}
\label{1d-ice}
The dipoles are initially assumed to interact via nearest neighbor dipole interactions, together with 
uniaxial anisotropy (along $\hat{y}$, parameter $K_1>0$) perpendicular to the chain and easy-plane anisotropy 
($xy$-plane, parameter $K_3>0$) as expected for thin islands of soft magnetic material on a substrate.
The 1D chain has $N$ dipoles $\mu {\bf S}_n$, where ${\bf S}_n$ are three-component dimensionless unit spin vectors separated by 
lattice constant $a$.  The nearest neighbor dipolar interaction constant is
\be
D = \frac{\mu_0 \mu^2}{4\pi\; a^3}.
\ee
The Hamiltonian for the chain is
\begin{align}
\label{Ham}
H = \sum_{n=1}^{N}  & \left\{ D\left[ {\bf S}_n\cdot {\bf S}_{n+1} -3({\bf S}_n\cdot \hat{x})({\bf S}_{n+1}\cdot \hat{x})\right] \right.
\nonumber \\
& \left. -K_1 \left(S_n^y\right)^2 +K_3 \left(S_n^z\right)^2 \right\}.
\end{align}
The dipolar interactions make neighboring dipole prefer to be perpendicular to the chain direction
and antiparallel to each other. The system can be analyzed as a two-sublattice problem. This is
most relevant for the $y$-alternating states.  Therefore, we introduce A and B sublattices,
for the odd and even sites, respectively. Let the spins now be labelled as ${\bf S}_n \rightarrow {\bf A}_n$,
and ${\bf S}_{n+1}\rightarrow {\bf B}_{n+1}$, for $n=1,3,5,...N$.  A two-spin cell becomes the basic unit in
the Hamiltonian.

\subsection{The uniform low energy states}

Initially, the static low-energy states are to be found.  Those have uniform aligned spins on each sublattice.
Let sublattice spin vectors {\bf A} and {\bf B} define the state.  A pair of neighboring sites has two bonds  and 
anisotropy terms for both sites.  The two-site Hamiltonian $H_{AB}$ is twice the energy per site $u$:
\begin{align}
\label{HAB}
H_{AB} = 2u = & 2D \left[ {\bf A}\cdot {\bf B} -3({\bf A}\cdot \hat{x})({\bf B}\cdot \hat{x})\right]
\nonumber \\
& -K_1\left(A_y^2+B_y^2\right) +K_3\left(A_z^2+B_z^2\right).
\end{align}
It is convenient to express the spins using out-of-plane and azimuthal angles $(\theta,\phi)$ as in
${\bf S}=(\cos\theta \cos\phi, \cos\theta \sin\phi, \sin\theta)$, which gives
\begin{align}
H_{AB} = & 2D\left[\sin\theta_A\sin\theta_B \right. \nonumber \\
& \left.+\cos\theta_A\cos\theta_B(-2\cos\phi_A\cos\phi_B+\sin\phi_A\sin\phi_B)\right]
\nonumber \\
& -K_1\left(\cos^2\!\theta_A \sin^2\!\phi_A+\cos^2\!\theta_B \sin^2\!\phi_B\right)
\nonumber \\
& +K_3\left(\sin^2\!\theta_A +\sin^2\!\theta_B\right)
\end{align}
Minimization of $H_{AB}$ with respect to the four angles leads to the possible states. The islands
are thin along the $z$-direction, which causes strong easy-plane ($K_3$) anisotropy so that static solutions
are planar, having $\theta_A=\theta_B=0$. Deviations of $\theta_A$ and $\theta_B$ away from zero are assumed 
in the stability analysis (Sec.\ \ref{E-stab}) and are present in the dynamic solutions (Sec.\ \ref{N-Modes}).  
There remains,
\begin{align}
\frac{\partial H_{AB}}{\partial \phi_A} = & 2\left\{ D(2\sin\phi_A\cos\phi_B+\cos\phi_A\sin\phi_B) \right.
\nonumber \\
& \left. -K_1 \sin\phi_A\cos\phi_A\right\} = 0,
\end{align}
together with the same relation with $A$ and $B$ interchanged.  These equations have three
physically distinct uniform-state solutions: (1) $x$-parallel states, with $\phi_A=\phi_B=0$ or $\phi_A=\phi_B=\pi$; 
(2) $y$-parallel states, with $\phi_A=\phi_B=\pm\frac{\pi}{2}$; and (3) $y$-alternating states,
with $\phi_A=-\phi_B=\pm\frac{\pi}{2}$. All are doubly degenerate.  The $x$-parallel states have
large anisotropy energy, while reducing their dipolar energy, while the $y$-alternating states 
tend to have low anisotropy energy and low dipolar energy.  The $y$-parallel states are intermediate;
their anisotropy energy is low but their dipolar energy is high.  

\begin{figure}
\includegraphics[width=\figwidth,angle=0]{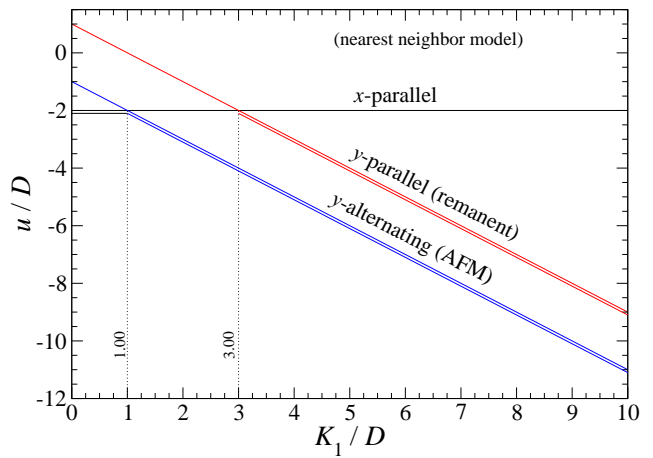}
\caption{\label{u123} The static per-site energy densities $u$ of the three planar metastable states, versus scaled
uniaxial anisotropy. The $x$-parallel state is stable only for $K_1<D$, $y$-alternating only for $K_1>D$, and 
$y$-parallel only for $K_1>3D$. Double (single) lines indicate local stability (instability) against weak perturbations.
Dynamic stability is shown to require $K_3>-D$ for all three states.}
\end{figure}

An indication of the stabilities of these states is obtained from the resulting energies per AB pair.
Consider planar states with $\theta_A=\theta_B=0$. 
%
%
For aligned sublattices ($\phi_A=\phi_B$), the two-site energy $H_{AB}(\phi_A,\phi_B)$ becomes 
\be
H_{AB}(\phi_A,\phi_A)= -4D+2(3D-K_1)\sin^2\!\phi_A.
\ee
This suggests that an $x$-parallel state (say, $\phi_A=\phi_B=0$) will be destabilized when $K_1>3D$,
because any small deviation in $\phi_A$ will lower the energy.  At the same time, it suggests that a $y$-parallel
state (say, $\phi_A=\phi_B=\frac{\pi}{2}$)  will be destabilized when $K_1<3D$.  A similar analysis
can be applied to states with antialigned sublattices ($\phi_A=-\phi_B$), where the two-site energy is
\be
H_{AB}(\phi_A,-\phi_A)= -4D+2(D-K_1)\sin^2\!\phi_A.
\ee
Considering small deviations around $\phi_A=0$, this shows the $x$-parallel states to be stable for $K_1<D$,
but unstable for $K_1>D$.  Considering instead deviations around a state with $\phi_A=\frac{\pi}{2}$ 
indicates that the $y$-alternating states will be unstable for $K_1<D$ but stable for $K_1>D$.  The per-site
energies $u(\phi_A,\phi_B)=\frac{1}{2}H_{AB}$ and stability requirements are summarized as follows:
\begin{align}
\label{states}
 \text{$x$-parallel: } & u(0,0)=-2D, &  K_1 < D,  \nonumber \\
 \text{$y$-parallel: } & u\left(\tfrac{\pi}{2},\tfrac{\pi}{2}\right)= -K_1+D, & K_1 > 3D, \nonumber \\
 \text{$y$-alternating: } & u\left(\tfrac{\pi}{2},-\tfrac{\pi}{2}\right)=-K_1-D, & K_1> D.
\end{align}
These estimates from the nearest neighbor model are plotted as functions of $K_1$ for fixed $D$ in Fig. \ref{u123}.
The $y$-parallel states become lower than $x$-parallel when they appear at $K_1> 3D$, and the $y$-alternating states
are lower than both $x$-parallel and $y$-parallel when they appear at $K_1>D$. The $y$-parallel states apparently are 
metastable, falling between the other two states for $K_1>3D$.  Starting from an alternating state, the system might 
arrive at a $y$-parallel state through the application of an external magnetic field along $y$, which is
then removed, making it a remanent-like state.  Such a state is expected to be locally stable against {\em small} 
perturbations even though lower energy states exist below it. 

%

\subsection{Planar state energetic stability?}
\label{E-stab}
The states found above must have a requirement on the easy-plane anisotropy ($K_3$) needed to maintain
local energetic stability (against weak perturbations).  It is important to consider small 
in-plane ($\phi_A,\phi_B$) and out-of-plane ($\theta_A,\theta_B$) deviations of the dipoles around each 
state to get a view of their local energetic stability. Let the in-plane angles become 
$\phi_A\rightarrow \bar{\phi}_A+\phi_A$ and  $\phi_B\rightarrow \bar{\phi}_B+\phi_B$, 
and the out-of-plane are $\theta_A, \theta_B$, where the overbar indicates the original state, and the other angles 
are the deviations.  In linear stability analysis, $H_{AB}$ is expanded to quadratic order in the deviations. 
The energies of in-plane and out-of-plane deviations separate, with 
\be
H_{AB} = \bar{H}+H_{\phi}+H_{\theta},
\ee
where $\bar{H}$ is the unperturbed state energy and $H_{\phi}$ and $H_{\theta}$ are the deviation energies.

\subsubsection{Stability of $x$-parallel states}
In the $x$-parallel state with $\bar{\phi}_A=\bar{\phi}_B=0$ and $\bar{H}=-4D$,  
the two-site deviation energies are
\begin{align}
H_{\phi} & =  (2D-K_1)\left(\phi_A^2+\phi_B^2\right)+2D\phi_A\phi_B, \nonumber \\
H_{\theta} & = \left(2D+K_3\right)\left(\theta_A^2+\theta_B^2\right)+2D\theta_A\theta_B.
\end{align}
These can be placed into matrix form, defining deviation vectors,
\be
\psi_{\theta}\equiv \left(\begin{array}{c} \theta_A \\ \theta_B \end{array} \right), \quad
\psi_{\phi}\equiv \left(\begin{array}{c} \phi_A \\ \phi_B \end{array} \right).
\ee
Then the deviation energies can be written as $H_{\phi}=\psi_{\phi}^{\dagger}M_{\phi} \psi_{\phi}$
and $H_{\theta} =  \psi_{\theta}^{\dagger}M_{\theta}\psi_{\theta}$, where there is a symmetric 
matrix for each part. The out-of-plane matrix is
\be
\label{Mxth}
M_{\theta} = \left( \begin{array}{cc} M_{\theta,0} & M_{\theta,1} \\ M_{\theta,1} & M_{\theta,0} \end{array} \right) 
=  \left( \begin{array}{cc} 2D+K_3 & D \\ D & 2D+K_3 \end{array} \right) 
\ee
Any instability of the $x$-parallel state will be exhibited as excitation of an eigenvector of $M_{\theta}$ with
a negative energy eigenvalue.  Considering the eigenvalue problem written as 
$M_{\theta} \psi_{\theta} = \sigma_{\theta} \psi_{\theta}$, the eigenvalues $\sigma_{\theta}$ are obtained by
\be
(M_{\theta,0}-\sigma_{\theta})^2-M_{\theta,1}^2=0 \ \implies \ \sigma_{\theta}^{\pm} = M_{\theta,0} \pm M_{\theta,1}.
\ee
There are symmetric and antisymmetric eigenvectors and their eigenvalues,
\be
\label{x-eigs}
\psi_{\theta}^{\pm} = \tfrac{1}{\sqrt{2}}(1,\pm 1),
\quad \sigma_{\theta}^{+}=3D+K_3,
\quad \sigma_{\theta}^{-}=D+K_3.
\ee
Both eigenvalues are real and positive, provided $K_3>-D$. If a deviation state is expressed as a linear combination, 
$\psi_{\theta}=c_{\theta}^{+} \psi_{\theta}^{+}+c_{\theta}^{-}\psi_{\theta}^{-}$, 
then the energy can only go upwards if any such deviation occurs:
\be
H_{\theta}   
= \left\vert c_{\theta}^{+}\right\vert^2 \sigma_{\theta}^{+}+\left\vert c_{\theta}^{-}\right\vert^2\sigma_{\theta}^{-}.
\ee
The $x$-parallel state is absolutely stable with respect to out-of-plane fluctuations for any $K_3>-D$. Even
when $K_3=0$, dipolar interactions alone provide a sufficient easy-plane anisotropy to guarantee
planar stability

For in-plane deviations, the matrix is
\be
\label{Mxph}
M_{\phi}=
\left( \begin{array}{cc} M_{\phi,0} & M_{\phi,1} \\ M_{\phi,1} & M_{\phi,0} \end{array} \right)
= \left( \begin{array}{cc} 2D-K_1 & D \\ D & 2D-K_1 \end{array} \right).
\ee
As the matrix has the same symmetric form as $M_{\theta}$, it has the same symmetric and antisymmetric eigenvectors,
but with different eigenvalues,
\be
\psi_{\phi}^{\pm} = \tfrac{1}{\sqrt{2}}(1,\pm 1),
\quad \sigma_{\phi}^{+} = 3D-K_1,
\quad \sigma_{\phi}^{-} = D-K_1.
\ee
Both eigenvalues remain positive as long as $K_1<D$, which confirms the stability requirement for in-plane deviations
found earlier for the $x$-parallel state.  The in-plane instability for $K_1>D$ is antisymmetric,
with out-of-phase sublattice deviations ($\phi_A=-\phi_B$). 

\subsubsection{Stability of $y$-parallel (remanent) states}
For the $y$-parallel states ($\bar{\phi}_A=\bar{\phi}_B=\pm\frac{\pi}{2}$) with unperturbed energy 
$\bar{H}=2(D-K_1)$, the deviation energies $H_{\theta}=\psi_{\theta}^{\dagger}M_{\theta}\psi_{\theta}$ and 
$H_{\phi}=\psi_{\phi}^{\dagger}M_{\phi}\psi_{\phi}$ are determined again by symmetric matrices of the
same form as encountered for $x$-parallel states, see Eqs.\ (\ref{Mxth}) and (\ref{Mxph}), but now the diagonal
and off-diagonal elements for the out-of-plane deviations are
\be
M_{\theta,0}=-D+K_1+K_3, \quad M_{\theta,1}=D.
\ee
The associated eigenvalues $\sigma_{\theta}^{\pm}=M_{\theta,0}\pm M_{\theta,1}$ are 
\be
\sigma_{\theta}^{+} = K_1+K_3, \quad \sigma_{\theta}^{-} = -2D +K_1+K_3.
\ee
Both eigenvalues stay positive and maintain stability if
\be
K_1+K_3 > 2D.
\ee
For in-plane deviations, the matrix elements are
\be
M_{\phi,0} = -D+K_1, \quad M_{\phi,1} = -2D.
\ee
The eigenvalues for symmetric or antisymmetric eigenvectors are
\be
\sigma_{\phi}^{+} = -3D+K_1, \quad \sigma_{\phi}^{-} =  D+K_1.
\ee
$\sigma_{\phi}^{-}$ is always positive as long as $K_1$ is positive, but $\sigma_{\phi}^{+}$ will
only stay positive and insure stability if $K_1>3D$.  This is more restrictive than the out-of-plane energy 
eigenvalues. Then $y$-parallel states are energetically stable only for $K_1>3D$ and any $K_3>-D$, 
although they are local minima or meta-stable states.

\subsubsection{Stability of $y$-alternating (AFM-ordered) states}
For the $y$-alternating state with $\bar{\phi}_A=\frac{\pi}{2}$ and $\bar{\phi}_B=-\frac{\pi}{2}$ and 
unperturbed energy $\bar{H}=-2(D+K_1)$, the same form of matrices again applies to the deviation energies.  
The diagonal and off-diagonal matrix elements for out-of-plane deviations are
\be
M_{\theta,0} = D+K_1+K_3, \quad M_{\theta,1} = D, 
\ee
which gives the eigenvalues,
\be
\sigma_{\theta}^{+}  = 2D+K_1+K_3, \quad \sigma_{\theta}^{-} = K_1+K_3.
\ee
These are always positive so there is no instability with respect to out-of-plane motions.
For in-plane deviations, the corresponding matrix elements are
\be
M_{\phi,0} = D+K_1, \quad M_{\phi,1} = 2D, 
\ee
which gives the eigenvalues that determine stability,
\be
\sigma_{\phi}^{+} = 3D+K_1, \quad \sigma_{\phi}^{-} = -D+K_1.
\ee
Here $\sigma_{\phi}^{+}$ is always positive, but $\sigma_{\phi}^{-}>0$ requires $K_1>D$ for stability against 
in-plane fluctuations.  One also sees $K_3>-D$ is the limiting condition for planar stability.

Therefore this energetic stability analysis confirms and expands upon the results of Eq.\ (\ref{states}).  At $K_1<D$, 
$x$-parallel states are the only stable ones.  With increasing $K_1$, the $y$-alternating states become stable
at $K_1 > D$, exactly at the point where the $x$-parallel states become destabilized.  Even more interesting
is that the $y$-parallel states become stable only for $K_1>3D$, when their energy falls {\em below} the already unstable
$x$-parallel states, but {\em above} the $y$-alternating states, see Fig.\ \ref{u123}, where local stability (instability)
is indicated by double (single) lines.  For $K_1>3D$, both $y$-parallel and $y$-alternating states are linearly stable 
against small perturbations. For planar stability, $K_3>-D$ is required for all three uniform states. Thus the
original assumption of easy-plane anisotropy ($K_3>0$) can be relaxed, as the dipolar interactions themselves produce
a weak easy-plane anisotropy.

\section{Normal mode oscillations around the states}
\label{N-Modes}
Now the linearized dynamics of the full chain is solved (beyond the two-site Hamiltonian).
It is shown that the dynamic oscillation mode frequencies are directly connected to the
energy eigenvalues above.  Any site has two nearest neighbors that exert torques on it 
(in the nearest neighbor approximation).  The magnetic dipole vectors are $\vec\mu_n=\mu {\bf S}_n$,
of magnitude $\mu$ with dimensionless spin vectors ${\bf S}_n$, and supposing a gyromagnetic ratio 
$\gamma$, they follow undamped dynamics from Hamiltonian (\ref{Ham}), using dot to indicate time
derivative \cite{Wysin15},
\be
\frac{1}{\gamma}\dot{\vec{\mu}}_n=\vec\mu_n\times \left(-\frac{\partial H}{\partial \vec\mu_n}\right).
\ee
A simplified way to write this is
\be
\label{Sdot}
\dot{\bf S}_n = {\bf S}_n\times {\bf F}_n,
\ee
where the Hamiltonian becomes $H=-\sum_n {\bf S}_n\cdot {\bf F}_n$ and the effective magnetic field on 
a site is ${\bf F}_n = -\partial H/\partial {\bf S}_n$, 
\be
{\bf F}_n = \kappa_1 S_n^y \hat{y} -\kappa_3 S_n^z \hat{z}+\delta_1 \sum_{k=n\pm 1} [3({\bf S}_k\cdot \hat{x})\hat{x}-{\bf S}_k].
\ee
The coupling parameters are
\be
\kappa_1\equiv \tfrac{2\gamma K_1}{\mu},
\quad \kappa_3\equiv \tfrac{2\gamma K_3}{\mu},
\quad \delta_1\equiv \tfrac{\gamma D}{\mu}.
\ee
The effective field components are
\begin{align}
F_n^x = &\ 2\delta_1 \left( S_{n-1}^x +S_{n+1}^x \right), \nonumber \\
F_n^y = & -\delta_1  \left( S_{n-1}^y +S_{n+1}^y \right) +\kappa_1 S_n^y, \nonumber \\
F_n^z = & -\delta_1  \left( S_{n-1}^z +S_{n+1}^z \right) -\kappa_3 S_n^z.
\end{align}
The oscillations take place relative to one of the three metastable states, whose 
unperturbed spin components are $\bar{\bf S}_{n}=\left(\bar{S}_{n}^x,\bar{S}_{n}^y,0\right)$, so that 
\be
{\bf S}_n = (\bar{S}_{n}^x+s_n^x, \bar{S}_{n}^y+s_n^y,s_n^z),
\ee
where ${\bf s}_n(t)=(s_n^x,s_n^y,s_n^z)$ are the small-amplitude time-dependent deviations.
These can also be expressed in terms of in-plane and out-of-plane angular deviations.  The equations of motion (\ref{Sdot})
are linearized in ${\bf s}_n$(t),  which leads to wave equations for the normal modes for each type of metastable state.

\subsection{Dynamics in the $x$-parallel states}
\begin{figure}
\includegraphics[width=\figwidth,angle=0]{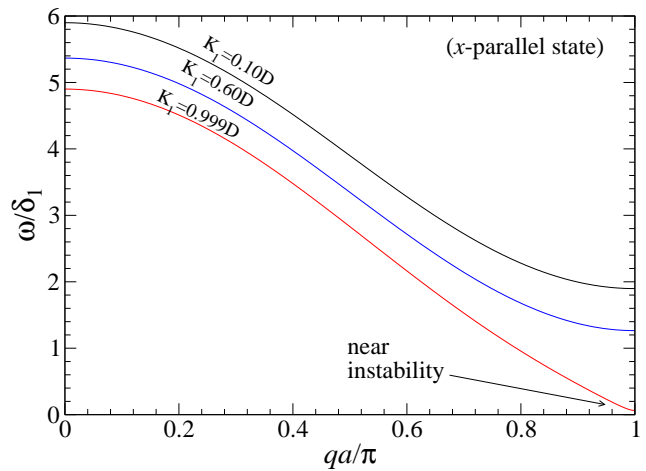}
\caption{\label{xparallel} The mode frequencies for a $x$-parallel state in the nearest neighbor model, with $K_3=0$, 
and indicated values of $K_1<D$.  The modes produce instability ($\omega\rightarrow 0$) at $qa=\pi$ as $K_1\rightarrow D$.}
\end{figure}

In an $x$-parallel state, $\bar{S}_{n}^x=1$ and $\bar{S}_{n}^y=0$ hold uniformly at all sites.  After linearization, 
the equations of motion imply $s_n^x=0$ and fixed spin length.  The in-plane and out-of-plane components satisfy
\begin{align}
\dot{s}_n^y = & +\delta_1 \left( 4s_n^z +s_{n-1}^z+s_{n+1}^z \right) + \kappa_3 s_n^z, \nonumber \\
\dot{s}_n^z = & -\delta_1 \left( 4s_n^y +s_{n-1}^y+s_{n+1}^y \right) + \kappa_1 s_n^y.
\end{align}
These are solved by traveling waves of the form $s_n^y(t)=s^y \exp[\ii(qna-\omega t)]$ and similar for the $z$-component, 
where $s^y$ is an amplitude,  $q$ is the wave vector, and $\omega$ is the frequency.  The resulting dispersion relation is
\be
\label{omx-dyn}
\frac{\omega }{ 4\delta_1} = \sqrt{\left(1 + \tfrac{1}{2}\cos qa -\tfrac{K_1}{2D}\right) 
\left(1 + \tfrac{1}{2}\cos qa +\tfrac{K_3}{2D}\right) }.
\ee
Examples are plotted in Fig.\ \ref{xparallel}, showing the maximum frequency at $q=0$, and a
gap at $qa=\pi$ that depends on the anisotropy. That gap goes to zero when instability occurs.
At the limit $qa=\pi$, both factors in the square root in (\ref{omx-dyn}) remain positive and imply real 
frequencies (and stability) of $x$--parallel states as long as 
\be
K_1 < D \quad \text{and} \quad K_3 > -D. 
\ee
To the contrary, an imaginary frequency signals instability if $K_1>D$ or $K_3<-D$. The latter condition means
that even when $K_3=0$ as used in Fig.\ \ref{xparallel} (i.e., no easy-plane anisotropy), the dipolar interactions 
work to stabilize a planar state of the dipoles. This holds for all three uniform states, see below.

These conditions on the anisotropy
parameters verify the earlier energetic stability analysis of $x$-parallel states.  In addition, the point where
instability first appears corresponds to $qa=\pi$ for $K_1=D$, with an excitation that is out-of-phase
on neighboring sites.  It agrees with the finding that the energy eigenvalue $\sigma_{\phi}^-=D-K_1$ associated with
the antisymmetric eigenvector $\psi_{\phi}^-$ becomes negative for $K_1>D$, see Eq.\ (\ref{x-eigs}).

\subsection{Dynamics in the $y$-parallel states}
\begin{figure}
\includegraphics[width=\figwidth,angle=0]{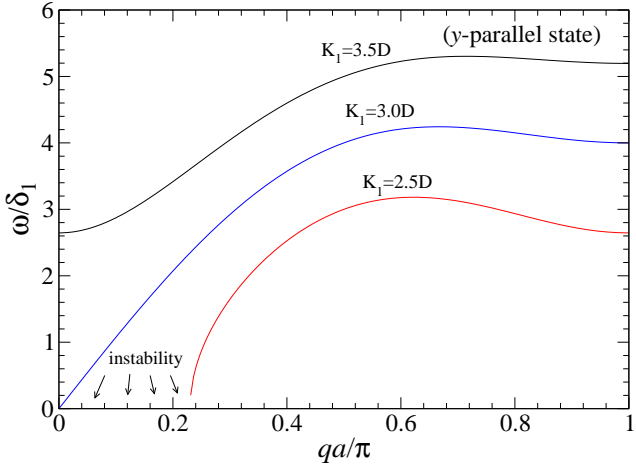}
\caption{\label{yparallel} The mode frequencies for a $y$-parallel state in the nearest neighbor model with $K_3=0$, and 
indicated uniaxial anisotropies $K_1$.  The states require $K_1>3D$ for real frequencies and stability.  The modes produce 
instability with $K_1<3D$, leading to imaginary $\omega$ near $qa=0$.}
\end{figure}

In a $y$-parallel state, $\bar{S}_{n}^x=0$ and $\bar{S}_{n}^y=1$ hold uniformly at all sites.  Now the deviation $s_n^y$ is zero
when the equations are linearized, and there remains oscillating in-plane and out-of-plane time-dependent deviations
that follow
\begin{align} 
\dot{s}_n^x = & +\delta_1  \left(2s_n^z-s_{n-1}^z-s_{n+1}^z\right) -\kappa_{13} s_n^z, 
\nonumber \\
\dot{s}_n^z = &   -2\delta_1 \left(s_n^x +s_{n-1}^x +s_{n+1}^x \right) +\kappa_1 s_n^x,
\end{align}
where the combined anisotropy constant is $\kappa_{13}\equiv \kappa_1+\kappa_3$.  
Traveling waves solve this system. Defining $K_{13}\equiv K_1+K_3$, the dispersion relation is
\be
\label{omy-dyn}
\frac{\omega}{2\delta_1} = \sqrt{ \left( 1 + 2\cos qa -\tfrac{K_1}{D} \right) 
\left(1-\cos qa -\tfrac{K_{13}}{D} \right) }.
\ee
The typical behavior of $\omega(q)$ is plotted in Fig.\ \ref{yparallel} for $K_3=0$ and various values of $K_1$.
The frequency must remain real for stability.  This requires both factors inside the square root
to be negative, such that their product is positive.  That leads to the conditions for stability of $y$-parallel states,
\be
K_1+K_3 > 2D \quad \text{and} \quad K_1>3D.
\ee
These imply the separated stability requirements,
\be
K_1>3D \quad \text{and} \quad K_3>-D.
\ee
The dipolar interactions by themselves already insure planar stability and easy-plane anisotropy is not essential.

The value of $K_{1}-3D$ controls the size of the gap at $q=0$.  On the other hand,
once $K_1<3D$ the uniaxial anisotropy is too weak to stabilize the $y$-parallel states.  This is 
the same stability limit as found in the energy eigenvalue analysis.  Also, the instability takes place now
at $qa=0$, where the second factor in the square root of the dispersion relation changes from negative to positive 
for $K_1<3D$.  This is reflected in the energy eigenvalue $\sigma_{\phi}^+=-3D+K_1$ becoming negative for $K_1<3D$, and
leading to in-phase in-plane deviations of the dipoles (i.e., $\psi_{\phi}^+$) as the excitation driving the 
instability.  Once $y$-parallel becomes unstable, the configuration should tend towards one of the stable 
$y$-alternating states.

\subsection{Dynamics in the $y$-alternating states}
\begin{figure}
\includegraphics[width=\figwidth,angle=0]{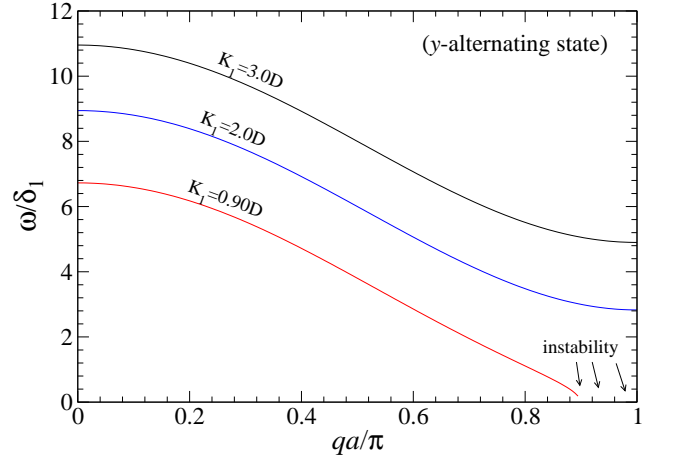}
\caption{\label{yalternating} The mode frequencies for a $y$-alternating state in the nearest neighbor model, with $K_3=0$, 
and $K_1>D$ needed for real frequencies and stability.  The modes produce instability (imaginary $\omega$)
for example with $K_1=0.90D$ near $qa=\pi$.}
\end{figure}

Finally consider the $y$-alternating state with $\bar{S}_{n}^x=\bar{S}_{n+1}^x=0$ and  
$\bar{S}_{n}^y=+1,\ \bar{S}_{n+1}^y=-1$, for $n=1,3,5...$  It is better to 
denote the spin deviations by sublattices, and
let $s_n^x\rightarrow a_n^x$ and $s_n^z\rightarrow a_n^z$ on the A-sublattice (odd sites $n$) and 
$s_{n+1}^x\rightarrow b_{n+1}^x$ with $s_{n+1}^z\rightarrow b_{n+1}^z$ on the B-sublattice (even sites $n+1$).
Any deviations in $y$-components are zero when the system is linearized.  The dynamic equations on
the A-sites are
\begin{align} 
\dot{a}_n^x = &   -\delta_1  \left( 2a_n^z +b_{n-1}^z+b_{n+1}^z\right) -\kappa_{13} a_n^z ,
\nonumber \\
\dot{a}_n^z = &   +2\delta_1 \left(a_n^x-b_{n-1}^x -b_{n+1}^x \right) +\kappa_1 a_n^x .
\end{align}
With the B-sites in an opposite unperturbed direction, their time derivative equations have reversed signs,
\begin{align} 
\dot{b}_n^x = &  +\delta_1  \left( 2b_n^z + a_{n-1}^z+a_{n+1}^z\right) +\kappa_{13} b_n^z,
\nonumber \\
\dot{b}_n^z = &   -2\delta_1 \left(b_n^x -a_{n-1}^x -a_{n+1}^x \right) -\kappa_1 b_n^x.
\end{align}
These equations are solved by traveling waves such as $a_n^x(t)=a^x \exp{[\ii(qna-\omega t)]}$ and
similar for the other components, however, it is necessary to form linear combinations of the fields on 
the two sublattices.  

Consider the linear combination that is antisymmetric in-plane but symmetric out-of-plane, defined by
\be
{\bf g}_n=(g_n^x,g_n^z) \equiv (a_n^x-b_n^x,\ a_n^z+b_n^z).
\ee
This produces a system in only this one field,
\begin{align}
\dot{g}_n^x = & -\delta_1 \left( 2 g_n^z+g_{n-1}^z+g_{n+1}^z\right)-\kappa_{13} g_n^z, 
\nonumber \\
\dot{g}_n^z = & +2\delta_1 \left( g_n^x+g_{n-1}^x+g_{n+1}^x\right)+\kappa_1 g_n^x.
\end{align}
The resulting dispersion relation for the ``g-modes'' is
\be
\label{gmodes-dyn}
\frac{\omega_g}{2\delta_1} = \sqrt{ \left(1 + 2\cos qa +\tfrac{K_1}{D}\right)
\left(1 + \cos qa +\tfrac{K_{13}}{D} \right) }.
\ee
This is plotted in Fig.\ \ref{yalternating} for $K_3=0$ and various values of $K_1$.
Now the frequency remains real provided both factors inside the square root remain positive. 
At the value $\cos qa=-1$, this leads to the conditions for stability, 
\be
K_1+K_3 >0 \quad \text{and} \quad K_1>D.
\ee
These imply $K_3 > -D$; dipolar interactions by themselves can provide planar stability.
The condition $K_1>D$ holds as long as the in-plane energy eigenvalue $\sigma_{\phi}^-=-D+K_1$ remains positive.  
Otherwise, when $K_1<D$, the modes of $y$-alternating states will be destabilized by an excitation at $qa=\pi$.
Due to the construction of the ${\bf g}$-field, that excitation will contain in-plane components that
are in-phase on the two sublattices, while the out-of-plane components will be out-of-phase.  The
destabilization will tend to drive the system into one of the (stable) $x$-parallel states.  The size
of the nonzero gap at $qa=\pi$ is determined by $K_1-D$.

Another linear combination that is symmetric in-plane but antisymmetric out-of-plane can be defined by
\be
{\bf h}_n=(h_n^x,h_n^z) \equiv (a_n^x+b_n^x,\ a_n^z-b_n^z).
\ee
This dynamics of this field is
\begin{align}
\dot{h}_n^x = & -\delta_1 \left( 2 h_n^z+h_{n-1}^z+h_{n+1}^z\right)-\kappa_{13} h_n^z, 
\nonumber \\
\dot{h}_n^z = & +2\delta_1 \left( h_n^x-h_{n-1}^x-h_{n+1}^x\right)+\kappa_1 h_n^x.
\end{align}
The ``h-modes'' here have the dispersion relation 
\be
\frac{\omega_h}{2\delta_1} = \sqrt{ \left(1 - \cos qa +\tfrac{K_{13}}{D} \right)
\left( 1 - 2\cos qa +\tfrac{K_1}{D} \right)}.
\ee
One can see that $\omega_h(q)=\omega_g(q+\frac{\pi}{a})$.  Therefore this set of solutions is already contained
in the g-modes, and leads to the same condition, $K_1>D$, for stability of the $y$-alternating states.

\section{Connecting the dynamic frequencies to the energy eigenvalues}
In classical mechanics, an out-of-plane spin component $S^z=\sin\theta$ is the momentum conjugate to the in-plane spin 
angle $\phi$ for a site.  In the two-sublattice model, the Hamilton equations of motion are obtained from the energy 
$H_{AB}$ by
\be
\frac{\mu}{\gamma}\frac{d}{dt}\phi = \frac{\partial H_{AB}}{\partial \sin\theta}, \hskip 0.4in
\frac{\mu}{\gamma}\frac{d}{dt}{\sin\theta} = -\frac{\partial H_{AB}}{\partial \phi}.
\ee
The energy $H_{AB}=2u$ assumes two uniform sublattices, and gives dynamics of a central A or central B site,
with two AB bonds contributing to the energy.  
The Hamiltonian is like that for a set of coupled oscillators, where squared $\theta$'s are kinetic energies and 
squared $\phi$'s are spring energies.
Considering small amplitude oscillations, $\sin\theta$ can be replaced by $\theta$ for each sublattice.  
Using states $\psi_{\phi}^{\dagger}\equiv (\phi_A,\phi_B)$ and $\psi_{\theta}^{\dagger} \equiv (\theta_A,\theta_B)$,  
the linearized matrix expressions for $H_{AB}=2u$ are of quadratic form,
\begin{align}
H_{AB} \approx \bar{H}+ &\psi_{\phi}^{\dagger}M_{\phi} \psi_{\phi}+\psi_{\theta}^{\dagger}M_{\theta} \psi_{\theta}  \\
 = \bar{H}+ &M_{\phi,0}\left(\phi_A^2+\phi_B^2\right)+2M_{\phi,1}\,\phi_A\phi_B
\nonumber \\
 +& M_{\theta,0}\left(\theta_A^2+\theta_B^2\right)+2M_{\theta,1}\,\theta_A\theta_B.  \nonumber
\end{align}
The matrix elements appearing here depend on the original state ($x$-parallel, etc., found earlier in \ref{E-stab}). 
The equations of motion \cite{Wysin15} are especially simple in matrix notation, 
\begin{align}
\dot{\psi}_{\phi}  = 2\tfrac{\gamma}{\mu}M_{\theta}\psi_{\theta}, 
&& \dot{\psi}_{\theta} = -2\tfrac{\gamma}{\mu}M_{\phi}\psi_{\phi}.
\end{align}
When combined they give separated eigenvalue problems,
\begin{align}
\ddot{\psi}_{\phi} & = -\omega^2 \psi_{\phi} = -\left(2\tfrac{\gamma}{\mu}\right)^2 M_{\theta}M_{\phi} \psi_{\phi}, \nonumber \\
\ddot{\psi}_{\theta} & = -\omega^2 \psi_{\theta} = -\left(2\tfrac{\gamma}{\mu}\right)^2 M_{\phi}M_{\theta} \psi_{\theta}.
\end{align}
These are identical eigenvalue problems, due to the symmetric structure of the matrices, and they determine the
eigenfrequencies.  We already know that the eigenvectors {\sl for both matrices} are $\psi^{\pm}=\frac{1}{\sqrt{2}}(1,\pm 1)$ 
and hence also for their products.  
That means the $\phi$ and $\theta$ oscillations have the same frequencies, determined by the eigenvalues $\sigma$ of the
$M_{\theta}$ and $M_{\phi}$ matrices,
\be
\omega^{\pm}  = 2\frac{\gamma}{\mu}\sqrt{\sigma_{\theta}^{\pm}\sigma_{\phi}^{\pm}}.
\ee
This necessarily requires that the $\theta$ and $\phi$ components are either {\sl both} in the $\psi^{+}$ eigenstate or {\sl both} in
the $\psi^{-}$ eigenstate.  It shows how the energy eigenvalues from $H_{AB}$ connect to the dynamics, and that the state is
unstable if the product $\sigma_{\theta}\sigma_{\phi}$ becomes negative.  When $\omega^{+}$ goes to zero, the instability
first appears at $qa=0$ (symmetrized sublattices), but if $\omega^{-}$ goes to zero, the instability is at $qa=\pi$
(antisymmetrized sublattices).

\subsection{Using the full chain Hamiltonian, nearest neighbor model}
The full $N$-site chain linearized dynamics comes from Hamiltonian (\ref{Ham})  approximated quadratically in
deviations $(\theta_n,\phi_n)$ around one of the stable states. Thus putting $\phi_n\rightarrow \bar{\phi}_n+\phi_n$, 
the Hamiltonian
\begin{align}
\label{Ham1}
H =  \sum_{n=1}^N & \Bigl[ D\left\{\sin\theta_n\sin\theta_{n+1} +\cos\theta_n\cos\theta_{n+1} \ \times \right.  \nonumber \\
& \left[-2\cos(\bar\phi_n+\phi_n)\cos(\bar\phi_{n+1}+\phi_{n+1}) \right. \nonumber \\
& \left.\left. +\sin(\bar\phi_n+\phi_n)\sin(\bar\phi_{n+1}+\phi_{n+1})\right] \right\}
\nonumber \\
& -K_1\cos^2\!\theta_n \sin^2(\bar\phi_n+\phi_n) +K_3\sin^2\!\theta_n  \Bigr]
\end{align}
is to be expanded.
The deviations are vectors of the in-plane and out-of-plane deviation angles,
\be
\psi_{\phi}^{\dagger}=(\phi_1,\phi_2,\phi_3,...\phi_N), \quad
\psi_{\theta}^{\dagger}=(\theta_1,\theta_2,\theta_3,...\theta_N). 
\ee
The system Hamiltonian has static state energy $\bar{H}$ plus quadratic contributions of deviations involving 
0$^{\rm th}$ and 1$^{\rm st}$ neighbors, 
$H \approx \bar{H} + \psi_{\phi}^{\dagger}M_{\phi} \psi_{\phi}+\psi_{\theta}^{\dagger}M_{\theta} \psi_{\theta}$, or
\begin{align}
\label{HN}
H \approx \bar{H} + \sum_n & \left[  M_{\phi,0} \phi_n^2 +2M_{\phi,1} \phi_n \phi_{n+1}  \right.  \nonumber \\
& \left. +M_{\theta,0}\theta_n^2+2M_{\theta,1}\theta_n\theta_{n+1} \right]
\end{align}
where the matrices are of tridiagonal form,
\be
\label{Mf}
M_{\phi} = \left( \begin{array}{ccccc}
M_{\phi,0} & M_{\phi,1} & 0          & 0          & ... \\
M_{\phi,1} & M_{\phi,0} & M_{\phi,1} & 0          & ...  \\
0          & M_{\phi,1} & M_{\phi,0} & M_{\phi,1} & ... \\
0          & 0          & M_{\phi,1} & M_{\phi,0} & ... \\
0          & 0          & 0          & M_{\phi,1} & ... \\
...        & ...        & ...        & ...        & ... \end{array} \right).
\ee
The diagonal ($M_{\phi,0}$) and nearest neighbor ($M_{\phi,1}$) matrix elements depend on the 
unperturbed state, see following subsections.
The linearized equations of motion from (\ref{HN}) are
\begin{align}
\label{motion}
\tfrac{\mu}{\gamma}\dot{\phi}_n & = +\tfrac{\partial H}{\partial\theta_n} 
= +2M_{\theta,0}\theta_n+2M_{\theta,1}(\theta_{n-1}+\theta_{n+1}), \nonumber \\
\tfrac{\mu}{\gamma}\dot{\theta}_n & = -\tfrac{\partial H}{\partial\phi_n} 
= -2M_{\phi,0}\phi_n-2M_{\phi,1}(\phi_{n-1}+\phi_{n+1}). 
\end{align}
These are very compact in matrix notation,
\begin{align}
\dot{\psi}_{\phi} & = 2\tfrac{\gamma}{\mu}M_{\theta}\psi_{\theta}, & \dot{\psi}_{\theta} = -2\tfrac{\gamma}{\mu}M_{\phi}\psi_{\phi}. 
\end{align}
Each matrix operator has eigenvalues, $\lambda_{\phi}^{(m)}, \lambda_{\theta}^{(m)}$, with
\be
\label{eigens}
M_{\phi} \psi_{\phi}^{(m)} = \lambda_{\phi}^{(m)} \psi_{\phi}^{(m)}, \quad
M_{\theta} \psi_{\theta}^{(m)} = \lambda_{\theta}^{(m)} \psi_{\theta}^{(m)},
\ee
where $(m)$ labels a simultaneous eigenvector of both $M_{\phi}$ and $M_{\theta}$.
Then the mode eigenfrequencies are
\be
\label{omegas}
\omega^{(m)} = 2\frac{\gamma}{\mu}\sqrt{\lambda_{\phi}^{(m)} \lambda_{\theta}^{(m)}}.
\ee
Due to the symmetric form of the matrices for this 1D problem, the dynamic mode
eigenvalues $\omega^{(m)}$ then are determined from the general energy eigenvalues for
deviations around the stable states.

\subsection{The eigenvalues for traveling wave solutions}
Consider traveling wave solutions and the eigenvalues of the $M$-matrices. With 
allowed wave vectors on a chain, $q=2\pi m/Na$, and a parameter $r\equiv e^{\ii qa}$, an 
assumed solution is
\be
\label{assumed}
\phi_n = \phi\, r^{n}, \quad \theta_n= \theta\, r^{n},
\ee
where $\phi$ and $\theta$ without subscripts are wave amplitudes at some origin $n=0$.
From (\ref{motion}), one row of the $M_{\phi}$ matrix acting on $\psi_{\phi}$ in the eigenvalue problem 
(\ref{eigens}) is 
\be
\lambda_{\phi}\, \phi r^n = \left[M_{\phi,0}+M_{\phi,1}(r^{-1}+r^{+1})\right]\, \phi r^n.
\ee
The eigenvalues of $M_{\phi}$ and $M_{\theta}$ easily result, 
\begin{align}
\lambda_{\phi}(q) & = M_{\phi,0}+2M_{\phi,1}\cos\, qa,  \nonumber \\
\lambda_{\theta}(q) & = M_{\theta,0}+2M_{\theta,1}\cos\, qa.
\end{align}
These eigenvalues then can be applied in expression (\ref{omegas}) and the dynamic mode frequencies
are determined in the nearest neighbor model, for each of the stable states, provided that $M_{\phi}$ and
$M_{\theta}$ have been determined.  The wave vector $q$ plays the role of the mode index $m$.
\vskip 0.1in

\subsubsection{Modes around $x$-parallel states}
For the $x$-parallel state with $\bar\phi_n=0$, the expansion of (\ref{Ham1}) leads to $\bar{H} = -2ND$ and
matrix elements, 
\begin{align}
\label{Mxpar}
M_{\phi,0} & =2D-K_1, & M_{\phi,1} & = \tfrac{1}{2}D, \nonumber \\
M_{\theta,0} & =2D+K_3, & M_{\theta,1} & = \tfrac{1}{2}D.
\end{align}
Then the $q$-dependent eigenvalues are
\begin{align}
\label{lams-xpar}
\lambda_{\phi}(q) &=  D(2+\cos qa)-K_1,
\nonumber \\
\lambda_{\theta}(q) &= D(2+\cos qa)+K_3.
\end{align}
Using (\ref{omegas}), the resulting mode frequencies are 
\be
\label{omx-e}
\frac{\omega(q)}{4\delta_1} = 
\sqrt{ (1+\tfrac{1}{2}\cos qa-\tfrac{K_1}{2D}) (1+\tfrac{1}{2}\cos qa+\tfrac{K_3}{2D}) },
\ee
which are the same as the earlier result (\ref{omx-dyn}) obtained from undamped dynamics.

\subsubsection{Modes around $y$-parallel states}
In a $y$-parallel state with $\bar\phi_n=\frac{\pi}{2}$, the expansion of the Hamiltonian (\ref{Ham1})
leads to $\bar{H}=N(D-K_1)$, while the matrix elements are
\begin{align}
\label{Mypar}
M_{\phi,0} & =-D+K_1, & M_{\phi,1} & = -D, \nonumber \\
M_{\theta,0} & =-D+K_1+K_3, & M_{\theta,1} & = \tfrac{1}{2}D.
\end{align}
That means the $q$-dependent eigenvalues are
\begin{align}
\lambda_{\phi}(q) &=  -D(1+2\cos qa)+K_1,
\nonumber \\
\lambda_{\theta}(q) &= D(-1+\cos qa)+K_1+K_3.
\end{align}
The $y$-parallel states require $K_1>3D$ for stability, so both of these eigenvalues are
positive for any allowed $q$.  Applying (\ref{omegas}), the dynamic mode frequencies are
\be
\frac{\omega(q)}{2\delta_1} = 
\sqrt{ \left(-1-2\cos qa+\tfrac{K_1}{D}\right) \left(-1+\cos qa+\tfrac{K_{13}}{D}\right) }.
\ee
This agrees with the earlier result (\ref{omy-dyn}), although here it has been written such that
both factors inside the square root are positive when $y$-parallel is stable.  

\subsubsection{Modes around $y$-alternating states}
For a $y$-alternating state with $\bar\phi_n=(-1)^n\frac{\pi}{2}$, the expansion of the Hamiltonian (\ref{Ham1})
gives $\bar{H}=-N(D+K_1)$, and the matrix elements are
\begin{align}
M_{\phi,0} & =D+K_1, & M_{\phi,1} & = D, \nonumber \\
M_{\theta,0} & =D+K_1+K_3, & M_{\theta,1} & = \tfrac{1}{2}D.
\end{align}
The resulting $q$-dependent eigenvalues are
\begin{align}
\lambda_{\phi}(q) &=  D(1+2\cos qa)+K_1,
\nonumber \\
\lambda_{\theta}(q) &= D(1+\cos qa)+K_1+K_3.
\end{align}
Both of these eigenvalues are positive for any $q$ as long as $K_1>D$, which is the previously
found stability requirement.  The dynamic mode frequencies become
\be
\frac{\omega(q)}{2\delta_1} = 
\sqrt{ \left(1+2\cos qa+\tfrac{K_1}{D}\right) \left(1+\cos qa+\tfrac{K_{13}}{D}\right) }.
\ee
That agrees with the earlier result (\ref{gmodes-dyn}) for the g-modes of a $y$-alternating state.

\section{Including long-range dipole interactions}
Long-range dipole interactions (LRD) can be included easily in the energy analysis.  The dipolar interaction strength 
in (\ref{Ham}) is $D$ at the nearest neighbor distance $a$.   This energy factor behaves as $1/r^3$.  So for 
2$^{\rm nd}$ nearest neighbors, the strength will be $D/{2^3}$; for 3$^{\rm rd}$ nearest neighbors it is $D/{3^3}$, 
and so on.  Along a 1D chain, the sum of all neighbors' energies to any distance can be constructed.

Only the first neighbor dipole terms were used in Eq.\ (\ref{Ham1}), as indicated in the subscripts $n+1$.
For the contribution of pairs of $k^{\rm th}$ neighbors (at separation distance $ka$), the dipolar part of the 
Hamiltonian is
\begin{align}
\label{Hamk}
H_k =  \tfrac{D}{k^3} \sum_{n=1}^N & \left\{\sin\theta_n\sin\theta_{n+k} +\cos\theta_n\cos\theta_{n+k} \ \times\right. \nonumber \\
& \left[-2\cos(\bar\phi_n+\phi_n)\cos(\bar\phi_{n+k}+\phi_{n+k}) \right. \nonumber \\
& \left.\left. +\sin(\bar\phi_n+\phi_n)\sin(\bar\phi_{n+k}+\phi_{n+k})\right] \right\}.
\end{align}
For $k^{\rm th}$ neighbors,  the zeroth, linear and quadratic terms in angular deviations are
obtained by expanding (\ref{Hamk}), which is most tractable when done separately for each meta-state.

\subsection{LRD in $x$-parallel states}
For $x$-parallel states ($\bar\phi_{n}=0$), expansion of (\ref{Hamk}) up to quadratic deviation terms 
for $k\ge 1$ gives
\begin{align}
H_k \approx \tfrac{D}{k^3} \sum_{n=1}^{N}
  & \left(-2+2\phi_n^2+\phi_n\phi_{n+k}+2\theta_n^2+\theta_n\theta_{n+k}\right).
\end{align}
The first term modifies the unperturbed $x$-parallel state energy by an amount $\bar{H}_k =-\frac{2}{k^3}ND$, and 
the others are quadratic interactions.  This implies shifts to the diagonal matrix elements [denoted $\Delta M_{\phi,0}^{(k)}$] 
and new elements ($M_{\phi,k}$) for $k^{\rm th}$ neighbor couplings,
\begin{align}
\Delta M_{\phi,0}^{(k)}& =\tfrac{2D}{k^3},
&  M_{\phi,k} & =  \tfrac{D}{2k^3}, \nonumber \\
\Delta M_{\theta,0}^{(k)}& =\tfrac{2D}{k^3}, 
&  M_{\theta,k} &  = \tfrac{D}{2k^3}. 
\end{align}
For example, keeping only 0$^{\rm th}$, 1$^{\rm st}$ and 2$^{\rm nd}$ nearest neighbors, a row of the eigenvalue problem, 
$M_{\phi}\psi_{\phi}=\lambda_{\phi}\psi_{\phi}$, is
\begin{align}
M_{\phi,0}\phi_n & +M_{\phi,1}(\phi_{n-1}+\phi_{n+1})  \\ 
& +M_{\phi,2}(\phi_{n-2}+\phi_{n+2}) = \lambda_{\phi}\phi_n, \nonumber
\end{align}
With the $x$-parallel matrix elements, this becomes
\begin{align}
\left(2D+\tfrac{1}{4}D-K_1\right)\phi_n & +\tfrac{1}{2}D(\phi_{n-1}+\phi_{n+1})  \\
& +\tfrac{1}{16}D(\phi_{n-2}+\phi_{n+2})=\lambda_{\phi}\phi_n. \nonumber
\end{align}
Using the traveling wave solution (\ref{assumed}) with $\phi_n\propto r^n$, the eigenvalue by
including up to 2$^{\rm nd}$ neighbors is
\be
\lambda_{\phi}  = -K_1+D\left(2+\cos qa\right)+\tfrac{1}{8}D\left(2+\cos 2qa\right).
\ee
The first two terms come from the nearest neighbor model, see (\ref{lams-xpar}).
The last term shows the effect of 2$^{\rm nd}$ neighbor interactions. It contains a shift 
by $\frac{1}{4}D$, as well a dependence on the doubled wave vector.

Third neighbors will have factors of $\frac{D}{3^3}$ and $\cos 3qa$, and so on for farther neighbors.
Assuming a chain of infinite length, the eigenvalue is a sum over increasingly distant
neighbor interactions, 
\begin{align}
\label{lfM}
\lambda_{\phi} & = \sum_{k=1}^{\infty} \left[ \Delta M_{\phi,0}^{(k)}+2M_{\phi,k}\cos kqa \right].
\end{align}
For $x$-parallel states this gives
\begin{align}
\lambda_{\phi}(q) & = -K_1+D\sum_{k=1}^{\infty} \frac{1}{k^3}\left(2+\cos kqa \right) 
\nonumber \\
& = -K_1+D\left[2\zeta(3)+{\rm Cl}_3(qa)\right].
\end{align}
This is written using the Riemann zeta function, whose needed value is $\zeta(3)\approx 1.20205$, 
combined with a Clausen function of order 3.

\begin{figure}
\includegraphics[width=\figwidth,angle=0]{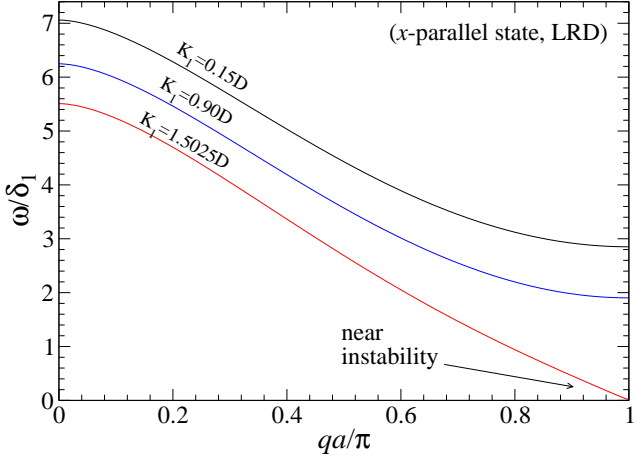}
\caption{\label{xpar-lrd} Mode frequencies for $x$-parallel states, including all long range dipole interactions,
with $K_3=0$.  The modes produce instability ($\omega$ acquires an imaginary part) near $qa=\pi$ as 
$K_1\rightarrow 1.502571\ D$.  The states have a greater range of stability than in the nearest neighbor model.}
\end{figure}

For the out-of-plane eigenvalue, the same procedure applies, but with $-K_1$ replaced by $+K_3$ (see Eq.\ \ref{Mxpar}), 
\begin{align}
\lambda_{\theta}(q)  & = K_3+D\sum_{k=1}^{\infty} \frac{1}{k^3}\left(2+\cos kqa \right) \nonumber \\
& = K_3+D\left[2\zeta(3)+{\rm Cl}_3(qa)\right].
\end{align}
According to (\ref{omegas}), they give the eigenfrequencies, 
\small
\be
\frac{\omega(q)}{4\delta_1} = \sqrt{\left[-\tfrac{K_1}{2D}+\zeta(3)+\tfrac{1}{2}{\rm Cl}_3(qa)\right]
\left[\tfrac{K_3}{2D}+\zeta(3)+\tfrac{1}{2}{\rm Cl}_3(qa)\right]}.
\ee
\normalsize
This is a simple modification of the frequencies in the nearest neighbor model, Eq.\ (\ref{omx-e}), but where 
${\rm Cl}_3(qa)$ replaces $\cos qa$, and $\zeta(3)$ has replaced a factor of 1.  Some typical dispersion relations
are shown in Fig.\ \ref{xpar-lrd}.  

Stability of the $x$-parallel state requires both eigenvalues to remain positive for any $q$. The most
negative value of the Clausen function occurs at $qa=\pi$, where $2\zeta(3)+{\rm Cl}_3(\pi)=1.502571...$
That means $\lambda_{\theta}$ is always positive, but $\lambda_{\phi}$ remains positive and the state is 
stable only as long as 
\be
K_1/D<1.502571.  
\ee
That is a 50\% increase compared to the nearest neighbor model. 


\subsection{LRD in $y$-parallel states}
\begin{figure}
\includegraphics[width=\figwidth,angle=0]{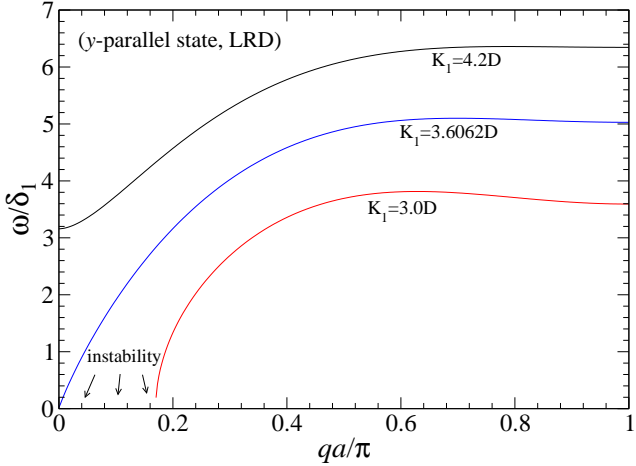}
\caption{\label{ypar-lrd} Mode frequencies for $y$-parallel states, including all long range dipole interactions,
with $K_3=0$.  The modes produce instability ($\omega$ has an imaginary part) near $qa=0$ as 
$K_1\rightarrow 3.60617\ D$ from above. For example, $y$-parallel is unstable for $K_1/D=3$, as
indicated by the presence of an imaginary frequency.}
\end{figure}

The same procedure can be applied to longer range dipolar interactions for $y$-parallel states,
such as the one with all $\bar\phi_n=\frac{\pi}{2}$.  Consider neighbors at separation distance $ka$.
From expansion of (\ref{Hamk}), the dipolar energy of $k^{\rm th}$-neighbor pairs to quadratic order in deviations is
\begin{align}
H_k \approx \tfrac{D}{k^3} \sum_{n=1}^{N} \left(1-\phi_n^2-2\phi_n\phi_{n+k}-\theta_n^2+\theta_n\theta_{n+k}\right).
\end{align}
The first term modifies the unperturbed $y$-parallel state energy by the amount $\bar{H}_k = \frac{1}{k^3}ND$.
The other parts determine the changes to diagonal matrix elements, and new elements for $k^{\rm th}$ 
neighbors,
\begin{align}
\Delta M_{\phi,0}^{(k)} & =-\tfrac{D}{k^3}, 
&  M_{\phi,k} &  = -\tfrac{D}{k^3}, \nonumber \\
\Delta M_{\theta,0}^{(k)} & =-\tfrac{D}{k^3}, 
&  M_{\theta,k} &  = +\tfrac{D}{2k^3}. 
\end{align}
As a specific example, when keeping up to $2^{\rm nd}$ neighbor dipole interactions, the in-plane eigenequation becomes
\begin{align}
\left(-D+K_1-\tfrac{1}{8}D\right)\phi_n &-D(\phi_{n-1}+\phi_{n+1})  \\
& -\tfrac{1}{8}D(\phi_{n-2}+\phi_{n+2})=\lambda_{\phi}\phi_n. \nonumber
\end{align}
The 3$^{\rm rd}$ neighbors will be similar but with a factor of $\frac{1}{27}$, and so on, so that
the result including  arbitrarily distant dipole terms is
\begin{align}
\lambda_{\phi}(q) & = K_1-D\sum_{k=1}^{\infty}\tfrac{1}{k^3}\left(1+2\cos kqa\right) \nonumber \\
& = K_1-D\left[\zeta(3)+2{\rm Cl}_3(qa)\right].
\end{align}
The same procedure applies to the out-of-plane system, but with $K_1$ replaced by $K_{13}$ and using 
$M_{\theta,1}=\frac{1}{2}D$ instead of $M_{\phi,1}=-D$, see Eq.\ (\ref{Mypar}).  The eigenvalues are 
\begin{align}
\lambda_{\theta}(q) &=K_1+K_3+D\sum_{k=1}^{\infty} \tfrac{1}{k^3}(-1+\cos kqa) \nonumber \\
& = K_{13}+D\left[-\zeta(3)+{\rm Cl}_3(qa)\right].
\end{align}
This gives the mode dispersion relation,
\small
\be
\frac{\omega(q)}{2\delta_1}  = \sqrt{\left[\tfrac{K_1}{D}-\zeta(3)-2{\rm Cl}_3(qa)]
[\tfrac{K_{13}}{D}-\zeta(3)+{\rm Cl}_3(qa)\right]}.
\ee
\normalsize
The dispersion relation is shown in Fig.\ \ref{ypar-lrd} for a few values of anisotropy.
For stability, both factors within the square root must be positive.  When $q=0$, one has ${\rm Cl}_3(0)=\zeta(3)$, 
which shows that stability requires 
\be
K_1/D> 3\zeta(3) \approx 3.606, 
\ee
which is a 20\% increase over the stability limit on $K_1$ in the nearest neighbor model.


\subsection{LRD in $y$-alternating states}
The analysis of $y$-alternating states, such as the one with $\bar\phi_n=(-1)^n\frac{\pi}{2}$, is different
from that for the parallel states, because $2^{\rm nd}$, $4^{\rm th}$, etc. neighbor spins are aligned, while 
$1^{\rm st}$, $3^{\rm rd}$, etc. neighbor spins are antialigned. 

In a $y$-alternating state, the in-plane factors in the dipolar energy $H_k$ [Eq.\ (\ref{Hamk})] 
for $k^{\rm th}$ neighbor pairs are
\small
\begin{align}
& \sin(\bar\phi_n+\phi_n)\sin(\bar\phi_{n+k}+\phi_{n+k}) \approx 
(-1)^k \left(1-\tfrac{1}{2}\phi_n^2-\tfrac{1}{2}\phi_{n+k}^2\right), \nonumber \\
& -2\cos(\bar\phi_n+\phi_n)\cos(\bar\phi_{n+k}+\phi_{n+k}) \approx (-1)^k \left(-2\phi_n \phi_{n+k}\right).  
\end{align}
\normalsize
The overall sign on these factors alternates with the separation, because a pair of dipoles perpendicular to the
chain will lower their energy by being antiparallel.  Then the dipolar energy (\ref{Hamk}) is approximated as
\small
\begin{align}
H_k & \approx \tfrac{D}{k^3} \sum_{n=1}^N \left[\theta_n\theta_{n+k} 
+\left(1-\tfrac{1}{2}\theta_n^2-\tfrac{1}{2}\theta_{n+k}^2\right)  \right. \\
& \left. \hskip 0.5in \times (-1)^k 
\left(1-\tfrac{1}{2}\phi_n^2-\tfrac{1}{2}\phi_{n+k}^2-2\phi_n\phi_{n+k}\right)\right]
\nonumber \\
& \approx \tfrac{D}{k^3} \sum_{n=1}^N 
\left[\theta_n\theta_{n+k}+(-1)^k \left(1-\phi_n^2-2\phi_n\phi_{n+k}-\theta_n^2\right)\right]. \nonumber
\end{align}
\normalsize
From that, the contribution to the energy of the $y$-alternating state is $\bar{H}_k = \frac{(-1)^k}{k^3}ND$.
The shifts in diagonal matrix elements and the elements for $k^{\rm th}$ neighbors are
\begin{align}
\Delta M_{\phi,0}^{(k)} & =-\tfrac{(-1)^k D}{k^3}, 
&  M_{\phi,k} & = -\tfrac{(-1)^k D}{k^3}, \nonumber \\
\Delta M_{\theta,0}^{(k)} & =-\tfrac{(-1)^k D}{k^3}, 
&  M_{\theta,k} & = \tfrac{D}{2k^3}. 
\end{align}
Then the sum in (\ref{lfM}) gives the in-plane eigenvalues, 
\begin{align}
\lambda_{\phi}(q) &= K_1-D\sum_{k=1}^{\infty} \tfrac{(-1)^k}{k^3} \left(1+2\cos kqa \right)
\end{align}
This alternating series is obtained from a Clausen function, at a shifted argument, {\it viz.}
\begin{align}
{\rm Cl}_3(qa+\pi) &= \sum_{k=1}^{\infty} \tfrac{1}{k^3}\cos[k(qa+\pi)] \\
& = \sum_{k=1}^{\infty} \tfrac{1}{k^3}\cos k\pi\, \cos kqa
= \sum_{k=1}^{\infty} \tfrac{(-1)^k}{k^3}\cos kqa . \nonumber
\end{align}
With the particular value, 
${\rm Cl}_3(\pi) = \sum_{n=1}^{\infty} \frac{(-1)^n}{n^3}  \approx -0.90154...$,
the eigenvalues are
\be
\lambda_{\phi}(q) = K_1-D\left[{\rm Cl}_3(\pi)+2 {\rm Cl}_3(qa+\pi)\right].
\ee

\begin{figure}
\includegraphics[width=\figwidth,angle=0]{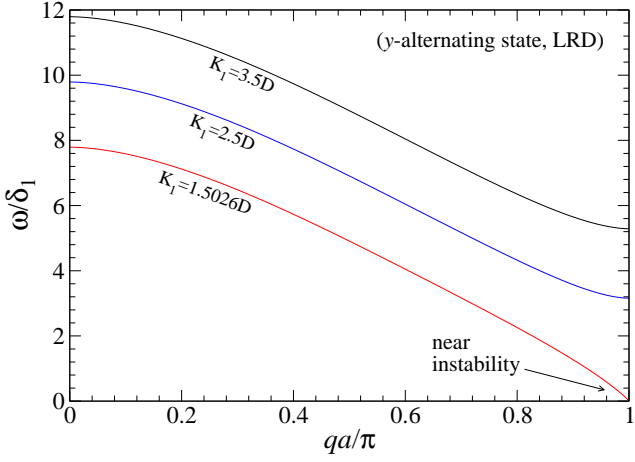}
\caption{\label{yalt-lrd} Mode frequencies for $y$-alternating states, including all long range dipole interactions,
with $K_3=0$.  The modes produce instability near $qa=\pi$ as $K_1\rightarrow 1.502571\ D$ from above.}
\end{figure}

For the out-of-plane eigenvalues, a sum like that in (\ref{lfM}) with the corresponding matrix
elements gives 
\begin{align}
\lambda_{\theta}(q) &= K_{13}-D\sum_{k=1}^{\infty} \tfrac{1}{k^3} \left((-1)^k+\cos kqa \right)
\nonumber \\
&= K_{13}+D\left[-{\rm Cl}_3(\pi)+{\rm Cl}_3(qa)\right].
\end{align}
Then the mode dispersion relation is
\small
\begin{align}
\frac{\omega(q)}{2\delta_1} & = & \left[\tfrac{K_1}{D}-{\rm Cl}_3(\pi)-2{\rm Cl}_3(qa+\pi)\right]^{1/2}
\nonumber \\ 
& & \times  \left[\tfrac{K_{13}}{D}-{\rm Cl}_3(\pi)+{\rm Cl}_3(qa)\right]^{1/2}.
\end{align}
\normalsize
Note how the factor Cl$_3(\pi)$ for the $y$-alternating states replaces Cl$_3(0)=\zeta(3)$ that appears
for $y$-parallel states, and there is the shifted argument of the Clausen function from the in-plane eigenvalue.
Example dispersion relations are shown in Fig.\ \ref{yalt-lrd}.
Regarding stability, $\lambda_{\theta}$ is always positive, so instability occurs when $\lambda_{\phi}$
becomes negative. Thus, the requirement on $K_1$ for stable $y$-alternating states is
\be
\tfrac{K_1}{D} > \left[{\rm Cl}_3(\pi)+2 {\rm Cl}_3(qa+\pi)\right]\vert_{qa\rightarrow\pi} \approx 1.502571...
\ee
This is the same anisotropy strength above which the $x$-parallel state goes unstable.

\begin{figure}
\includegraphics[width=\figwidth,angle=0]{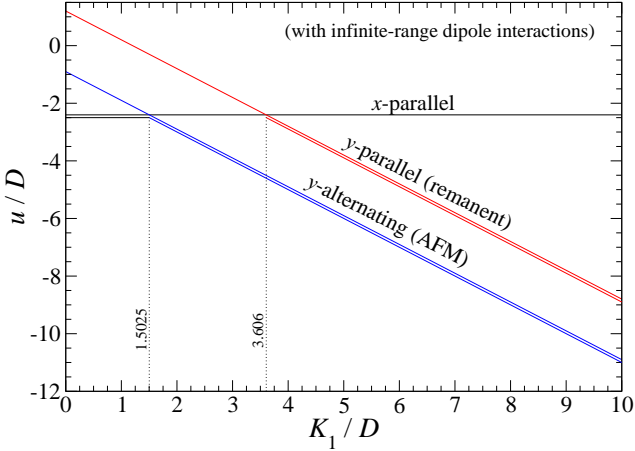}
\caption{\label{u123-lrd} The static per-site energies of the three types of states, 
when all long-range dipole interactions are included. The $x$-parallel states are now stable only for $K_1< 1.50257\, D$,
$y$-alternating for $K_1> 1.50257\, D$,  and $y$-parallel for $K_1>3.606\, D$. Double (single) lines indicate local 
stability (instability) against weak perturbations.}
\end{figure}

\subsection{State energies with all dipole interactions}
It is important also to get the per-site energy, $u=H/N$, including all long range dipole interactions.  
That involves combining the $\bar{H}_k$ contributions and anisotropy contributions.
For $x$-parallel states, the sum  is,
\be
u = \sum_{k=1}^{\infty} \frac{-2D}{k^3} = -2D\zeta(3) \approx -2.404\, D.
\ee
That is $-0.404\, D$ lower than that from the nearest neighbor model.
Along with the wider range of $K_1$ that insures stability, this shows that long range dipole 
interactions help to stabilize the $x$-parallel states.

For $y$-parallel states, the per-site energy is
\begin{align}
u & = -K_1+\sum_{k=1}^{\infty} \frac{D}{k^3} 
\nonumber \\ 
& = -K_1+D\zeta(3) \approx -K_1+1.202\, D.
\end{align}
That is $0.202\, D$ higher than found for the nearest neighbor model.  Viewed together with the modified 
stability requirement, this indicates that long range dipole interactions slightly reduce the stability of 
$y$-parallel states.

For $y$-alternating states, the energy per site becomes
\begin{align}
u & = -K_1+\sum_{k=1}^{\infty} \frac{(-1)^k D}{k^3} \nonumber \\
& = -K_1+D {\rm Cl}_3(\pi) \approx -K_1-0.9015\, D.
\end{align}
That is $0.0985\, D$ higher than found in the nearest neighbor model, because 2$^{\rm nd}$ neighbor dipoles
are antialigned and have high dipolar energy.

The energies of all three states are summarized graphically in Fig.\ \ref{u123-lrd}.  
This analysis indicates that $y$-parallel and $y$-alternating states are both locally stable in the regime where
$K_1/D > 3\zeta(3)$, even though $y$-alternating states have lower energy.
%

\section{Discussion \& Conclusions}
An array of elongated magnetic islands arranged as in Fig.\ \ref{1d-islands} has been shown to have three uniform 
states, whose linear stability depends on the uniaxial anisotropy strength $K_1$ relative to the nearest neighbor
dipolar interaction strength $D$. 
The conclusions are reached both by looking at the behavior of the states' energy-related eigenvalues $\lambda_{\phi}$
and $\lambda_{\theta}$, and the related frequencies of the linearized oscillations about the states.

Including infinite range dipole interactions, the states' linear stability regimes, shown in Fig.\ \ref{u123-lrd}
with double lines,  are summarized as follows.
At very weak anisotropy ($K_1<1.50257\, D$), the $x$-parallel states will be the only stable states, where the 
dipoles minimize their dipolar energy with little cost in anisotropy energy.
For an intermediate range of anisotropy ($1.50257\, D < K_1 < 3.606\, D$), the AFM-ordered $y$-alternating states
have the lowest energy and become the only stable states.  Both the dipolar and anisotropy energies
are minimized. 
For strong anisotropy ($3.606\, D < K_1$), the $y$-alternating states still have the lowest energy
and are stable. 
The remanent FM-ordered $y$-parallel states are locally stable against small perturbations (they have only positive 
$\lambda_{\phi}$ and $\lambda_{\theta}$ eigenvalues), although they are higher in energy than the $y$-alternating 
states by $2.103\, D$.  
They would be metastable with respect to strong perturbations.
In the $y$-parallel remanent states, the anisotropy energy is minimized and dominates greatly over the 
non-optimal dipolar energy. 

The remanent states could be produced experimentally by application of a magnetic field transverse to the 
chain, which is slowly turned off. 
The $y$-alternating states could be obtained from a remanent state either by application of a field along the
chain ($x$) direction, which is then reduced to zero, or, by a field opposite to the original direction
of magnetization.
Other switching between states might be implemented with the help of AC and rotating demagnetization protocols 
as used in square and rectangular lattice artificial magnetic ice \cite{Ke08,Nisoli10,Ribeiro+17} to reach its 
various states.

This type of system might be the basis for new detectors or devices, and allows for various ways to control
the design.  
Although a static island array will have fixed values of $K_1$ and $D$, one can imagine these might be
modulated, for example, by applying pressure \cite{Edberg21} to an elastic substrate or other modified new materials. 
If the system is designed
near one of the critical anisotropy points ($K_1 \approx 1.50257\, D$ or $K_1\approx 3.606\, D$), it might
be possible to induced a strain in the medium that shifts the stability point between two of the states.
Alternatively, application of magnetic fields may be sufficient to control switching between the $y$-alternating
and $y$-parallel states.  
These effects invite further investigation.

\end{document}